\documentstyle[multicol,pre,aps,epsf]{revtex}
\begin{document}

\title{Spin glasses without time-reversal symmetry and the absence of a 
genuine
structural glass transition}

\author{Barbara Drossel}
\address{School of Physics and Astronomy, Raymond and 
Beverley Sackler Faculty of Exact Sciences, Tel Aviv 
University, Tel Aviv 69978, Israel}
\author{Hemant Bokil\footnote{Present Address: Program in Neuroscience,
University of Maryland, Baltimore MD 21201, U.S.A.}}
\address{Abdus Salam ICTP, Strada Costiera 11, 34100 Trieste, 
Italy}
\author{M. A. Moore}
\address{Department of Physics, University of Manchester, 
Manchester M13 9PL, U.K.}
\date{\today} 
\maketitle

\begin{abstract}
  We study the three-spin model and the Ising spin glass in a field
  using Migdal-Kadanoff approximation. The flows of the couplings and
  fields indicate no phase transition, but they show even for the
  three-spin model a slow crossover to the asymptotic high-temperature
  behaviour for strong values of the couplings. We also evaluated a
  quantity that is a measure of the degree of non-self-averaging, and
  we found that it can become large for certain ranges of the
  parameters and the system sizes. For the spin glass in a field the
  maximum of non-self-averaging follows for given system size a line
  that resembles the de Almeida-Thouless line. We conclude that
  non-self-averaging found in Monte-Carlo simulations cannot be taken
  as evidence for the existence of a low-temperature phase with
  replica-symmetry breaking. Models similar to the three-spin model
  have been extensively discussed in order to provide a description of
  structural glasses. Their theory at mean-field level resembles the
  mode-coupling theory of real glasses. At that level the one-step
  replica symmetry approach breaking predicts two transitions, the
  first transition being dynamical and the second thermodynamical.
  Our results suggest that in real finite dimensional glasses there
  will be no genuine transitions at all, but that some features of
  mean-field theory could still provide some useful insights.
\end{abstract}

\begin{multicols}{2}
  
\section{Introduction}
\label{intro}

Despite over two decades of work, the controversy concerning the
nature of the ordered phase of short-range Ising spin glasses
continues.  Monte-Carlo simulations of three- and four-dimensional
systems appear to be providing evidence for replica symmetry breaking
(RSB) in these systems (for recent reviews see \cite{marinari,mprrz}).
However, recent developments have cast doubt on this interpretation of
the Monte Carlo data.  In a series of papers on the Ising spin glass
within the Migdal-Kadanoff approximation (MKA), we showed that the
equilibrium Monte Carlo data in three and four dimensions that had
been interpreted in the past as giving evidence for RSB can actually
be interpreted quite easily within the droplet picture, with apparent
RSB effects being attributed to a crossover between critical behaviour
and the asymptotic droplet-like behaviour for small system
sizes\cite{us1,us2,us3,us4,us5}.  We also showed that system sizes
well beyond the reach of current simulations would probably be
required in order to unambiguously see droplet-like behaviour.  Very
recently, a third view on the nature of the low-temperature phase if
spin glasses has emerged. In this picture, there exist droplet
excitations in short scales, but on large scales there are system-wide
excitations that cost only a finite energy in the thermodynamic limit,
and that have a surface whose fractal dimension is less than the space
dimension \cite{hm,bou,km,py1,py2}. It remains to be seen whether
these excitations survive at larger system sizes. Within MKA, fractal
excitations are not possible, and the signatures of these excitations
found in Monte-Carlo simulations are therefore not present in MKA. 

There is a close connection between the question of the nature of the
spin glass phase and that of the existence of a phase transition in a
spin glass in an external field. Mean-field theory predicts a phase
transition to a spin glass phase with RSB along the so-called de
Almeida-Thouless line. The droplet picture predicts no transition. The
reason is that in the presence of a field time reversal symmetry is
broken, and there is no symmetry left that could possibly be broken by
a phase transition, except for replica symmetry. Monte-Carlo
simulation of the spin glass in a field \cite{mpz98,mnz98,mnzppr98}
show some evidence of a phase transition, and in particular of
non-self-averaging, however, the situation is complicated by the
presence of large finite-size effects due to crossover phenomena. For
this reason, Parisi et al \cite{ppr99} studied a different system that
also has broken time-reversal symmetry, but is expected not to have
strong crossover effects. This system is the three-spin model, where
the two-spin products in the Ising spin glass without field are
replaced by three-spin products. The numerical evidence for a phase
transition in the four-dimensional system seems good, if a quantity
that measures the degree of non-self-averaging is studied.

It is the purpose of this paper to study the two mentioned 
systems without
time-reversal symmetry in MKA, in order to check whether a 
similar
degree of non-self-averaging could be produced by finite-size 
effects.
The flow of the coupling constants shows that for both systems 
there
exists only one attractive fixed point which corresponds to a
paramagnet in a random field, and that there is consequently 
no phase
transition within MKA. Nevertheless, in both systems the 
coupling
constants increase initially for sufficiently low 
temperatures,
indicating that for small system sizes there might be the 
appearance
of a phase transition. Then, we looked at the 
non-self-averaging
parameter in both models within MKA, for various system sizes 
and
parameter values. We found a behaviour similar to the one 
reported for
the Monte-Carlo simulations, and apparent RSB for system sizes 
similar
to theirs. Furthermore, for the spin glass in a field, the 
maximum of
the non-self-averaging parameter as function of the field (for 
fixed
system size) marks a line that can be interpreted as a remnant 
of the
de Almeida-Thouless line. 

Some insights into what might be expected in the finite dimensional
three-spin model can be obtained from the mean-field solution of the
p-spin model (for a review of which see \cite{review}). It has an
analytical solution in the limit where $p$ goes to infinity which can
be obtained by a one-step replica symmetry breaking scheme. For our
purposes the solution is best understood in terms of metastable states
which can be identified with the solutions of Thouless-Anderson-Palmer
(TAP)-like equations \cite{kirk}.  The partition function is obtained
from the integral
\begin{equation}
Z =  \int^{f_{max}}_{f_{min}}df\exp(N\sigma(f)-Nf\beta),
\end{equation}
where $\beta=1/k_BT$ and $N$ is the number of spins in the system. $f$
is the free energy density of a TAP state. Solutions of the TAP
equations exist for $f_{min}<f<f_{max}$, and the number of solutions
at free energy density $f$ is exponentially large and given by
$\exp(N\sigma(f))$, with $\sigma(f)$ vanishing at both $f_{min}$ and
$f_{max}$. For temperatures in the interval $0<T<T_S$ the integral is
dominated as $N\rightarrow\infty$ by the lower limit of the integral,
i.e. states whose free energy $Nf$ differs from the state of lowest
energy by only a finite amount. For temperatures in the interval
$T_S<T<T_D$ the integral can again be done by steepest descents and is
dominated by some value of $f$ lying in the interval
$f_{min}<f<f_{max}$. As the temperature approaches $T_D$ this value
tends to $f_{max}$. In the temperature interval $T_S<T<T_D$ an
exponentially large number of states contribute to the thermodynamics
in contrast to the situation below $T_S$ where only a finite number
contribute.  Above $T_D$, only the trivial paramagnetic state
contributes. There are thus two phase transitions at mean-field level.
The lower temperature transition at $T_S$ is accompanied by
singularities in the free energy but at the higher temperature
transition $T_D$ the free energy is smooth and the presence of a
transition is best inferred from singularities in the dynamics.

Now for a finite dimensional system metastable states with $f>f_{min}$
are unstable. (Imagine in such a state converting a block of spins of
linear dimension $L$ to have the orientations that they would have in
the lowest state; the free energy gain will be or order $L^d$; the
energy cost of creating such a region will be no more than the cost of
breaking all the bonds at the surface of the region, $L^{d-1}$. Thus
the possibility of nucleating lower free energy states prevents the
existence of metastable states in finite dimensional systems). As the
transition at the higher temperature $T_D$ involves the metastable
states only, one deduces that it will not exist in a finite
dimensional system. The only transition which could possibly survive
to finite dimensions is the one associated with $T_S$. Our studies
however indicate that it too probably does not occur in finite
dimensional systems.
 
Our study of the three-spin model had another, perhaps physically more
significant motivation. The three-spin model and its cousins have been
extensively studied at mean-field level as models of structural
glasses \cite{kirk2}. The higher temperature transition, $T_D$, whose
signature is purely dynamical (and very similar to that in the
mode-coupling approach to real glasses \cite{Gotze}), while the
transition at the lower temperature, $T_S$, is associated the Kauzmann
temperature, $T_K$ \cite{Kauzmann}, the temperature at which the
configurational entropy of the glass goes to zero.  It has been a
common belief of many workers for several decades that there is no
genuine transition at $T_K$. Recently this belief has been strongly
reinforced by the novel Monte Carlo simulation of Santen and Krauth
\cite{santen}, who found no evidence of a genuine transition in a
simulation in which the particles could be properly equilibriated. Our
work strengthens the argument that no genuine transition analogous to
$T_S$ or $T_K$ will exist in finite dimensions.  Nevertheless we can
see echoes of the mean-field results in our calculations. It is our
belief that future work should focus on how the singularities in
mean-field results are "rounded-off" in finite dimensions.

This paper is organized as follows: First, we  define the 
models and quantities studied in this paper. Then, we   
describe the MKA for
the two models studied in this paper. We  use two different
methods in order to make sure that the results do not depend 
on the
particular implementation of the MKA. Next, we  study the
three-spin model and give our results for the flow of the 
coupling
constants, and for the non-self-averaging parameter. In 
section V,
we  discuss the spin glass in an external field. Again, we 
give results for the flows and the degree of 
non-self-averaging. We
 also give scaling arguments based on the droplet picture that
explain most of our findings. Finally, we  summarize and 
discuss
our results.

\section{Models and Definitions}
\label{definitions}

The Edwards-Anderson spin-glass Hamiltonian ${\cal H}$ in the 
presence of a
uniform external magnetic field $h$ is given by
\begin{equation}
-\beta {\cal H}(\sigma)=\sum_{\langle i,j\rangle} J_{ij} 
\sigma_i \sigma_j + h\sum_i \sigma_i,
\end{equation}
where the Ising spins $\sigma_i$ can take the values $\pm 1$, 
and the
nearest-neighbor couplings $J_{ij}$ are independent from each 
other
and Gaussian distributed with a standard deviation $J$. For
convenience, the Boltzmann-factor $\beta$ is absorbed 
into the
couplings and fields. Without a field $h$, the model has a
low-temperature phase with nonvanishing correlations $\langle 
S_i S_j \rangle $ even for spins that are far
apart. According to the droplet picture, this phase is unique 
(up to a
global flip of all spins), and it is destroyed as soon as the 
field is
turned on. The reason is that the field induces regions of a
sufficiently large radius to flip if the magnetization of this 
region
opposes the field.  The radius $r$ of these regions is 
obtained from
the condition that the gain in magnetic energy, $hr^{d/2}$ 
becomes
comparable to the loss in coupling energy, $Jr^{\theta}$, 
leading to 
\begin{equation}
r \sim (J/h)^{1/(d/2-\theta)}.
\label{droplet}
\end{equation}
 Here, $d$ is the dimension of the system,
and $\theta$ the scaling dimension of domain walls. Beyond the 
 radius $r$,
the long-range correlations of the spin-glass phase are 
destroyed. 

In contrast, the RSB picture predicts the existence of 
infinitely many
different phases of comparable free energy in the absence of a 
field.
With increasing field $h$, the number of phases decreases, and 
it
becomes one at the de Almeida-Thouless line $h_c(J)$. At the 
critical
spin-glass transition $J_c$, the critical field $h_c$ 
vanishes, and it
diverges to infinity as the coupling strength $J$ diverges 
(i.e., as
the temperature goes to zero). 

The numerical analysis of the spin glass in a field is 
hampered by
strong crossover effects. Crossover effects are expected to be 
less
strong in the three-spin model, because it has no tunable 
parameter
that can restore time reversal symmetry and lead to strong 
crossover
effects when small \cite{mnzppr98}.  In the most easily 
tractable
version of this model, each site is occupied by two Ising 
spins,
$\sigma_i^{(1)}$ and $\sigma_i^{(2)}$, and the Hamiltonian is 
given by

\begin{eqnarray}
-\beta{\cal H}(\sigma) &=& \sum_{ij} \large(J_{ij}^{(1)}
 \sigma_i^{(1)}\sigma^{(2)}_i\sigma_j^{(1)} +J_{ij}^{(2)}
 \sigma_i^{(1)}\sigma^{(2)}_i\sigma^{(2)}_j \nonumber\\ 
&+&J_{ij}^{(3)}
 \sigma_i^{(1)}\sigma_j^{(1)}\sigma^{(2)}_j +J_{ij}^{(4)} 
\sigma^{(2)}_i
\sigma_j^{(1)}\sigma^{(2)}_j\large)\, ,
\end{eqnarray}
where ${ij}$ are nearest-neighbor pairs, and the couplings
$J_{ij}^{(n)}$ are chosen independently from a Gaussian 
distribution
with zero mean and width $J$. When the signs of all spins are
reversed, the sign of the Hamiltonian changes also, indicating 
the
violation of time-reversal symmetry. 

If finite-dimensional systems have no RSB, this model has no
phase transition since there is no symmetry that could be 
broken.  On
the other hand, if RSB occurs in finite-dimensional spin 
glasses, the
three-spin model could show a phase transition at some 
critical
coupling strength $J_c$.

It has proven useful to consider two identical copies 
(replicas) of
the system, with the spins $\{ \sigma_i\}$ and $\{\tau_i\}$, 
and to measure overlaps between them. This gives
information about the structure of the low-temperature phase, 
in
particular about the number of pure states.
The main quantity studied in this paper is the parameter $A$ 
which measures the degree of non-self averaging, and is 
defined by 
\begin{equation}
A = \frac{\left[\langle (q-\langle q \rangle)^2 
\rangle^2\right]}{\left[\langle (q-\langle q \rangle)^2 
\rangle \right]^2} - 1,
\label{A}
\end{equation}
where $\langle ...\rangle$ and $\left[...\right]$ denote the
thermodynamic and disorder average respectively.  The overlap 
$q$
between the two replicas is given by
$$q=\frac{1}{N} \sum_i \sigma_i \tau_i$$
for the Ising spin glass in a
field, and by
$$q=\frac{1}{2N} \sum_i (\sigma_i^{(1)} \tau_i^{(1)} 
+\sigma_i^{(2)}
\tau_i^{(2)} )$$
for the three-spin model. $N$ is the number of sites
in the system.

$A$ is most easily evaluated by introducing a coupling between 
the two replicas, and by differentiating with respect to it. 
The Hamiltonian for the coupled system is then
$$-\beta{\cal H}_\epsilon(\sigma,\tau) = -\beta{\cal 
H}(\sigma)-\beta{\cal H}(\tau)+\epsilon N q.$$ The mean 
overlap is given by the expression
$$\langle q \rangle = \left[\frac{1}{N} 
\frac{\partial}{\partial \epsilon} \ln Z\right]_{\epsilon = 
0},$$
and its variance by
$$\langle (q-\langle q \rangle)^2\rangle = \left[\frac{1}{N} 
\frac{\partial^2}{\partial \epsilon^2} \ln Z\right]_{\epsilon 
= 0}\ . $$
$Z$ is the partition function. 

In systems with RSB, the probability distribution $P(q)$ of $q$ is
broad, and $A$ has a nonzero limit in the limit of infinite system
size. On the other hand, in the absence of RSB, each sample has only
one, sample-independent value of $q$, and $A$ vanishes in the
thermodynamic limit.  Consequently, an $A$ that increases with
increasing system size could be taken as an indicator of RSB. However,
we will see in this paper that even systems without RSB can show an
increasing $A$ over a wide range of system sizes and parameters.

\section{The Migdal Kadanoff Approximation}
\label{MKA}

\begin{figure}
\centerline{
\epsfysize=0.3\columnwidth{\epsfbox{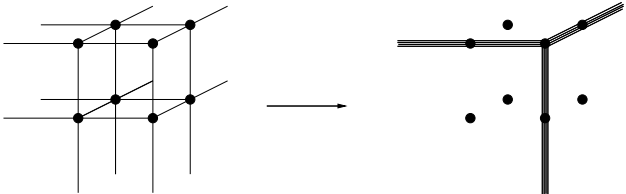}}}
\narrowtext{\caption{The Migdal-Kadanoff bond-moving scheme 
for a cubic lattice.}
\label{bm} }
\end{figure}

The Migdal-Kadanoff approximation (MKA) is a real-space
renormalization group that gives approximate recursion 
relations for
the various coupling constants. An exact decimation, which 
consists in
taking the trace over all those spins that do not belong to 
the
coarse-grained lattice, generates higher-order couplings 
between spins
of more that two sites, and is therefore untractable. In order 
to
circumvent this problem, the MKA moves the bonds of a 
hypercubic
lattice before each decimation step in such a way that no 
higher-order
couplings can be generated. If the bond-moving shall be 
symmetric with
respect to the different space directions, one ends up with 
the scheme
represented in Figure \ref{bm}.
In a $d$-dimensional lattice, $2^{d-1}$ bonds are superimposed 
as a
consequence of bond-moving. In the absence of field terms 
(i.e. of
terms that couple only spins that sit on the same site $i$, 
and are
therefore not in a clear way associated with bonds), the 
$2^{d-1}$
coupling constants of each of the $d$ bundles of bonds per
coarse-grained unit cell simply add up, and the ``naked'' 
spins that
are left behind have no couplings.  Taking the trace over the 
$d$
spins that sit on the $d$ main bonds leads to coarse-grained 
coupling
constants between neighboring spins on the coarse-grained 
lattice.
Taking the trace over the naked spins, gives only a constant
contribution to the partition function, which can be 
neglected.  This
decimation procedure is iterated $n$ times on a lattice of 
linear size
$L=2^n$, until a single unit cell is left over. Assuming 
periodic
boundary conditions, one can then take the trace over the 
final spin.

\begin{figure}
\centerline{
\epsfysize=0.2\columnwidth{\epsfbox{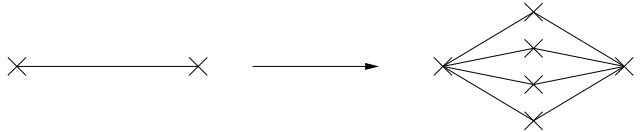}}
}
\narrowtext{\caption{Construction of a hierarchical 
lattice.}\label{fig1}}
\end{figure}

The flow of the coupling constants in this scheme results from
alternating the addition of $2^{d-1}$ bonds with linking two 
of these
new bonds together and taking the trace over the middle spin.
Essentially the same flow results when a decimation is done on 
a
hierarchical lattice that is constructed iteratively by 
replacing each
bond by $2^d$ bonds, as indicated in Fig.~\ref{fig1}. The 
total number
of bonds after $n$ iterations is $2^{dn}$. Spin decimation on 
such a
lattice is done by taking the trace over the spins that are 
highest on
this hierarchy (i.e., that were added last during the 
construction
procedure). At each decimation step, first the trace is taken 
over the
middle spin of two linked bonds, and then $2^{d-1}$ bonds are 
added
together to form a new bond, until the lowest level is reached 
and the
trace over the remaining two spins is calculated 
\cite{southern77}.
Apart from the fact that the order of bond-adding and 
decimation is
reversed, the recursion relations for the coupling constants 
are
obtained by the same procedure as for the bond-moving 
algorithm.

This equivalence between a bond-moving procedure for a 
hypercubic
lattice and a hierarchical lattice hold no longer when field 
terms are
present. In both of our models, these field terms are present 
either
from the beginning (Ising spin glass in a field), or they are
initially absent, but are generated during the decimation 
procedure
(three-spin model). For the bond-moving procedure, one has to 
decide
whether the fields shall also be moved (and to where), or 
whether they
shall remain at their sites, or whether part of them shall be 
moved.
This creates a certain freedom in the renormalization scheme, 
and the
most plausible choice is determined by the requirement that 
the flows
of the fields near the zero-temperature fixed point and at the
infinite-temperature fixed point shall be those of the 
hypercubic
lattice. However, if the main results of the MKA shall be 
generic,
they should not depend on the precise implementation of the
bond-moving algorithm.  Otherwise one might doubt that the MKA
reflects the features of the real system. For this reason, we 
have
performed the MKA for a variety of different implementations, 
and we
found our main conclusions concering $A$ to be independent of 
the
implementation. In the following sections, we will give our 
results
for the hierarchical lattice and for a bond-moving scheme that 
has the
correct flow of the field terms.

The treatment of a hierarchical lattice in the presence of 
field terms
is straightforward. In order to understand that the flows of 
the
fields are different on a hierarchical lattice compared to a
hypercubic lattice, let us consider a situation where the 
flows of the
couplings go to zero with increasing iteration number, which 
is the
situation that we will encounter below for both models. As 
long as the
couplings are nonzero, each decimation step generates a 
contribution
to the fields at the sites that are left over. The two corner 
spins,
which are left over until the end, receive consequently 
$2^{n(d-1)}$
field contributions from the first decimation, 
$2^{(n-1)(d-1)}$ from
the second iteration, and so on, until the couplings are 
virtually
zero. For twice the system size, i.e., for a lattice with 
$n+1$
levels, the mean of the field contribution to the corner spins 
due to
decimations is larger by a factor $2^{(d-1)}$, and so is the 
variance of
the field contribution. Even though the couplings go to zero 
after a
certain number of iterations, the fields keep growing. 

In contrast, on a hypercubic lattice, the fields must remain 
constant
as soon as the couplings have become zero. Clearly, this can 
only be
achieved if fields terms are not moved to the sites that will 
not be
traced over.  On the other hand, field terms must not be left 
with the
``naked'' spins. The reason is that near the zero-temperature 
fixed
point where the couplings are very large, all fields must add 
up under
renormalization. Fields must therefore always stay with spins 
that are
coupled to other spins. For this reason, fields should be 
moved to
those $d$ sites that sit at the middle of the main bonds.  
Even with
this restriction, there remains some freedom in choosing which 
field
should be moved where. In our simulations, we treated field 
terms as
belonging to bonds. The initial fields were evenly distributed 
between
the ends of all bonds, and the fields generated during 
decimations
naturally end up at the ends of bonds. When a bond was moved, 
we moved
all its field terms to that end that was to be traced over 
next.

For the Ising spin glass in a field, the recursion of four 
different
parameters must be considered when studying the flow diagram 
and
thermodynamic quantities. These are the two-spin coupling, the 
two
fields on the two ends of a bond, and a constant. If one 
evaluates
quantities related to the overlap between two replicas, each 
site has
two spins, leading to 16 parameters. The same number of 
parameters
occurs for the three-spin model, if only one replica is 
needed, as,
e.g., for the flows and the phase diagram. For the evaluation 
of $A$,
we need two replicas, leading 256 couplings. Luckily, the 
decimation
step can treat all 256 parameters with the same formula, which
involves a $256 \times 256$ matrix that is calculated once at 
the
beginning of the program.

\section{The Three-Spin Model}

This model was studied using Monte-Carlo simulations in four
dimensions in \cite{mnzppr98,mnzppr99,ppr99}, and evidence for RSB was
found.  The authors of \cite{mnzppr99,ppr99} found in particular that
the non-self-averaging parameter $A$ is small for large temperatures,
and becomes large for smaller temperatures. Curves for different
system size $L=3,4,5,6$ cross nearly at the same temperature, and
below this temperature $A$ increases with increasing $L$. Thus, the
degree of non-self averaging increases with the system size, just as
can be expected for a replica-symmetry breaking transition.
Monte-Carlo simulations \cite{mnzppr98,mnzppr99,ppr99} are usually
done with couplings $J=\pm 1$. The precise distribution of the
couplings should however not affect the universality class.

Analytical results were obtained for the $p$-spin model in
mean-field theory, where one-step RSB was found. This means that the
ground state has a nonzero probability of being occupied below a
critical temperature $T_S$ (see Section \ref{intro}). This mean-field
scenario is fundamentally different from the full RSB claimed to be
seen in Monte-Carlo simulations of the four-dimensional system. Thus,
the argument usually employed for spin glasses that mean-field like
behaviour can be found in finite-dimensional short-range systems,
fails here.

\begin{figure}
  \centerline{
    \epsfysize=0.7\columnwidth{\epsfbox{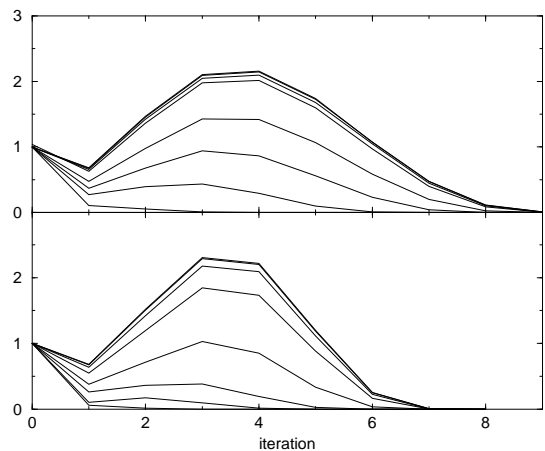}}}
  \narrowtext{\caption{Flow of the three-spin coupling strength $J$
      for bond-moving (top) and the hierarchical lattice (bottom) for
      $d=4$, divided by the initial coupling strength. The curves
      correspond to $J=0.065, 0.13, 0.18, 0.26, 0.5552, 1,2,5$
      (bond-moving) and $J=0.13, 0.26, 0.35, 0.52, 1, 2, 5, 10$
      (hierarchical lattice) from bottom to top.  }\label{flowj3} }
\end{figure}

In the following, we show using MKA that the assumption of the absence
of crossover effects in this model is incorrect, and that $A$ might at
low temperatures and for small system sizes increase with increasing
system size even if the system is self-averaging in the thermodynamic
limit. We mainly focus on the case of four dimensions, but report also
some results in d=2 and 3. Let us first discuss the flow of the
coupling constants as the system is renormalized. Because each bond is
connected to 4 spins, the flow of 16 coupling constants has to be
considered.  In order to obtain this flow, we iterated the recursion
relation on a set of $10^6$ bonds.  At each iteration, each of the new
set of $10^6$ bonds was generated by randomly choosing $2^d$ bonds
from the old set. For a hierarchical lattice, where the generated
fields remain at that end of a bond at which they are generated, we
first took the trace over the inner spins of each of the $2^{d-1}$
pairs of bonds, and than we added the resulting bonds; for the
bond-moving procedure described in the previous section, we first
generated two bunches of $2^{d-1}$ bonds each, then moved the fields
of all but the ``original'' bond of each bunch to the inner spin, and
took the trace over the inner spin.  Figure \ref{flowj3} shows the
flow of the width of the three-spin couplings for different initial
width in four dimensions, for the two different algorithms.
One can see that for weak coupling (or, equivalently, high
temperature) the coupling strength decreases quickly with 
increasing
system size. However, for stronger coupling or lower 
temperatures, the
coupling strength increases during the first few iterations, 
and
decreases afterwards. The maximum is reached between the 3rd 
and 4th
iteration, or between $L=8$ and $L=16$. For sufficiently 
strong
coupling, the curves reach an asymptotic shape. On the 
hierarchical
lattice, where the fields grow without bounds, the 3-spin 
couplings
decrease to zero faster than with bond-moving. Furthermore, 
curves for
the hierarchical lattice seem to correspond roughly to those 
of the
bond-moving procedure if the three-spin couplings are divided 
by a
number around 3. The reason is that the first step during the
bond-moving procedure summarizes 8 bonds to one new bond. The 
width of
the three-spin coupling is therefore increased by a factor of
$\sqrt{8}$ in four dimensions. In order to compare to the 
hierarchical
lattice or to Monte-Carlo simulations on a hypercubic lattice, 
one
should divide the coupling strength of the bond-moving 
procedure by
$\sqrt{8}$. 

If one considered only systems of sizes up to 8, one would
get the illusion of a phase transition with a $(1/J)_c$ around 
3 or 4,
a value which is not far from the one given for $T_c$ in 
\cite{ppr99}.
(Note that these authors have kept the coupling strength fixed 
at $\pm
1$, and varied the temperature. Their $T$ corresponds 
therefore to our
$1/J$.)

\begin{figure}
  \centerline{    
\epsfysize=0.7\columnwidth{\epsfbox{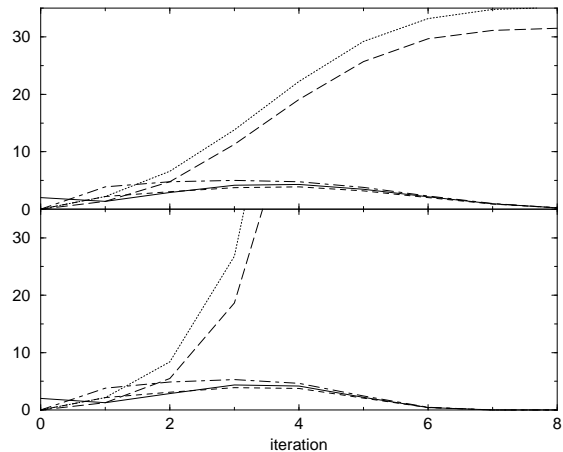
}}}
  \narrowtext{\caption{Flow of the width of the field (long 
dashed),
      of the on-site two-spin coupling (dotted), the two-spin 
coupling
      across a link (dashed), the three-spin coupling (solid), 
and the
      four-spin coupling (dot-dashed) for bondmoving (top) and 
on the
      hierarchical lattice (bottom) for $d=4$, for an initial
      three-spin coupling $J=2$.  }\label{flow_j3ini=2} }
\end{figure}

Figure \ref{flow_j3ini=2} shows the flow of the widths of the 
different coupling constants for an initial width of the 
three-spin coupling $J=2$. 
One can see that the strengths of the field and of the on-site 
two-spin
coupling (which can also be viewed as a ``field'') increase 
rapidly
and without limits for the hierarchical lattice, and that they 
saturate at
a finite value in the bond-moving case. The two- three- and 
four-spin
couplings increase during the first few iterations, and then 
decrease
again.  Thus, our three-spin model corresponds on large scales 
to a
system only with random fields and random couplings between 
the
$\sigma$ and $\tau$ spins on the same site. There are no 
couplings
between spins on different sites on large enough scales, but 
sites
are independent from each other. Only on small scales could 
one get
the impression that the system has long-range correlations.
However, these system sizes are exactly the ones studied in
\cite{mnzppr98,mnzppr99,ppr99}.

\begin{figure}
  \centerline{
\epsfysize=0.7\columnwidth{\epsfbox{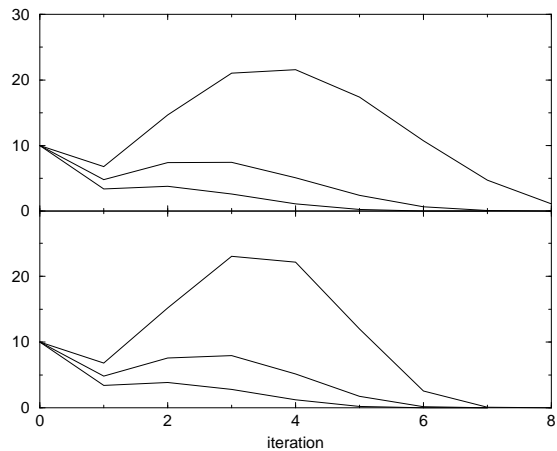}}}
  \narrowtext{\caption{Flow of the width of the field for an 
initial value $J=10$ in $d=2,3,4$ dimensions (from bottom to 
top). The top graph is for bondmoving, the bottom graph for a 
hierarchical lattice.
}\label{flow_j3=10} }
\end{figure}

The crossover regime becomes larger with increasing dimension. 
Figure
\ref{flow_j3=10} shows the flow of the three-spin coupling for 
an
initial value $J=10$ in $d=2,3,4$ dimensions.  Clearly, the 
strength
of the increase and the range of system sizes over which this 
increase
occurs increases with increasing dimension. One can therefore 
expect
that in even higher dimensions, the apparent phase transition 
becomes more pronounced. 

\begin{figure}
  \centerline{
    \epsfysize=0.7\columnwidth{\epsfbox{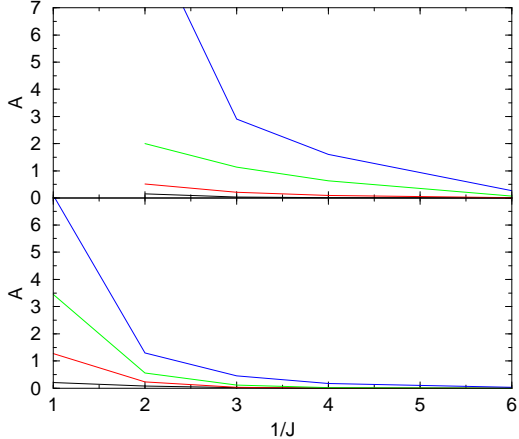}}}
  \narrowtext{\caption{The non-self-averaging parameter $A$ for 
$L=2,4,8,16$ (from bottom to top), and $d=4$. The average is 
taken over $10000$ samples for the smallest system size and 
$100$ samples for the largest. The top graph is again for 
bond-moving, a
nd the bottom graph on the hierarchical lattice.
}\label{a_d=4} }
\end{figure}

Next, let us study the non-self-averaging parameter $A$.  As 
explained
in the previous section, $A$ can be evaluated by introducing a
coupling between two identical replicas of the system. Since 
there are
now 8 spins associated with each bond, the number of couplings 
that
have to be evaluated in MKA approximation is $2^8=256$.
Figure \ref{a_d=4} shows $A$ as function of the coupling 
strength for different system sizes up to 16, in 4 dimensions. 
Larger system sizes could not be studied due to limitations in
computer time. One can see that $A$ increases with increasing 
system
size whereever it is appreciably different from zero, and 
reaches
large values. This figure gives the impression that the system 
shows
non self-averaging.  Of course, for larger system sizes, $A$ 
must
eventually decrease again since we know that there is 
self-averaging
in the thermodynamic limit.  In contrast to the Monte-Carlo 
simulation
results \cite{mnzppr99,ppr99}, our curves for $A$ do not 
intersect at
a coupling strength and $A$ value of the order 1.  We have 
performed a
similar simulation in $d=2$ dimensions and found that $A$ 
increases as
the system size increases over the range $L=2,4,8,16$. 
However, for
$L=32$ and $L=64$, $A$ decreases. If we assume that the system 
size
for which $A$ is largest increases with each dimension by a 
factor 2,
as it does for the flow of the couplings, we can expect that 
in $d=4$
the system sizes for which a decrease in $A$ can be seen is 
beyond
$L=64$.

To summarize this section, we have shown that the three-spin 
model,
even in situations where we know that it self-averages in the
thermodynamic limit, can show indications of 
non-self-averaging at
those system sizes typically studied in simulations. Evidence 
for
non-self-averaging found in Monte-Carlo simulations must 
therefore be
taken with caution as it might be misleading.

\section{Ising Spin Glass in a Magnetic Field}

Monte-Carlo simulations in four dimensions show some 
indication of RSB
\cite{mpz98,mnz98,mnzppr98}. Just as for the three-spin model 
and for
the spin glass without external field, these findings may 
again be due
to finite-size effects and to the closeness to the critical
temperature. Indeed, an investigation of the ground-state 
structure of
a spin glass in a magnetic field \cite{hm99} shows no 
indication of
RSB. (See, however, the discussion in \cite{comment,reply}.)

In order to test for finite-size effects, we studied
the spin glass in a field using MKA, and determined the
non-self-averaging parameter $A$ as function of the system 
parameters.
We found that the degree of non-self-averaging can be large 
for the
system sizes typically used in simulations, in particular when 
the
contribution of the field to the free energy is comparable to 
that of
the couplings. While most published Monte-Carlo simulations 
were done
in four dimensions, we chose to study the MKA in three 
dimensions, in
order to be able to go to larger system sizes. Because there 
are three
parameters to be varied (the system size, the field, and the 
two-spin
couplings), many data points had to be collected, and this is 
done
faster in 3 dimensions. Of course, we expect that the results 
of the
MKA are similar in four dimensions, if the exponents for 3
dimensions are replaced with those for 4 dimensions. Just as 
for the
three-spin model, the apparent non-self-averaging should 
become even
stronger in 4 dimensions.

\begin{figure}
  \centerline{
\epsfysize=0.7\columnwidth{\epsfbox{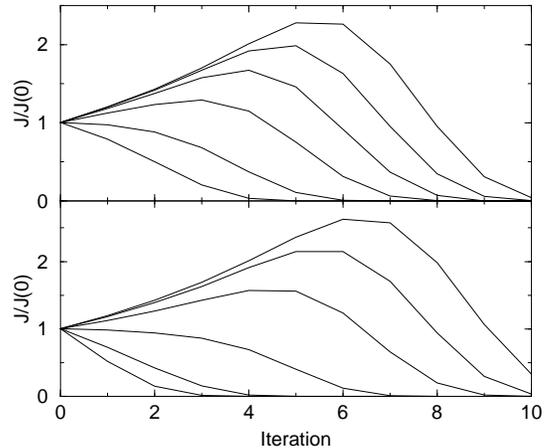}}}
  \narrowtext{\caption{Flow of the two-spin coupling strength 
$J$,
      divided by the initial strenggth, for bond-moving (top) 
and the
      hierarchical lattice (bottom) for $d=3$, divided by the 
initial
      coupling strength. The curves correspond to $J=0.3, 0.5, 
1, 2,
      4, 8 $ from bottom to top, for both graphs. The data 
where
      obtained from a ``cloud'' of 50000 bonds.  
}\label{flows} }
\end{figure}

First, let us study the flows of the couplings and fields. The
decimation procedure leads to the creation of random fields, 
while the
mean value of the field is not changed. Figure \ref{flows} 
shows the
flow of the two-spin coupling $J$ for various initial values, 
and for
a fixed field $h=0.1$. For initial couplings larger than the 
critical
coupling (1.13 for the hierarchical lattice and 0.55 for 
bondmoving),
the coupling strength decreases immediately.  However, if the 
initial
coupling strength is sufficiently deep in the low-temperature 
phase,
it increases first, until the random field has become strong 
enough to
have a reducing effect on the coupling strength.  Ultimately, 
the flow
goes to a fixed point where the coupling strength is zero. On 
the
hierarchical lattice, the width of the field keeps growing
indefinitely, while it saturates in the bond-moving case, as 
discussed
in section \ref{MKA}. Clearly, there is no phase transition in 
the
presence of an external field, but there are strong crossover 
effects
if the field is small.

\begin{figure}
  \centerline{
\epsfysize=0.7\columnwidth{\epsfbox{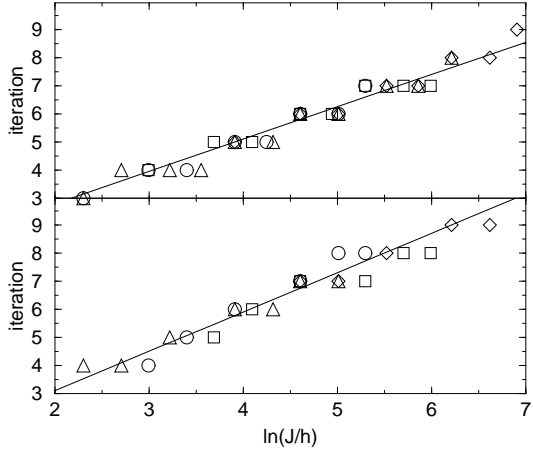}}}
  \narrowtext{\caption{Iteration number for which the two-spin
      coupling $J$ reaches its maximum, as function of 
$\ln(J/h)$, for
      different values of $h$. The symbols stand for $h=0.2$
      (triangle), $h=0.1$ (circle), $h=0.05$ (square), and 
$h=0.02$
      (diamond). The lines have the slope 1.15 and offset $C=0.5$ 
(bond-moving, top graph), and slope 1.4 and offset $C=0.3$ 
(hierarchical lattice, bottom graph).  }\label{flowmax} }
\end{figure}

As mentioned in Section \ref{definitions}, the droplet picture
predicts that beyond a length scale $r \sim 
(J/h)^{1/(d/2-\theta)}$ the
contributions of the field and of the couplings to the free 
energy
become comparable, and we expect that the strength of the 
couplings
decreases beyond this scale. In order to test this prediction, 
we have
plotted in Figure \ref{flowmax} the iteration number for which 
the
two-spin coupling is largest versus the logarithm of $J/h$. 
It should follow a law 
$$\log_2 L = \frac{1}{\ln 2 ((d/2)-\theta)} \ln(J/h) + C \simeq 1.15
\ln(J/h) + C\, ,$$
with a suitable constant $C$. As the figure shows,
the data for bond-moving agree nicely with this prediction.  for the
hierarchical lattice, the slope is larger and has a value around 1.4.
This might be due to the fact that the field increases faster on the
hierarchical lattice, leading to an earlier reduction in the coupling
strength than predicted by the scaling theory.

\begin{figure}
  \centerline{
\epsfysize=0.7\columnwidth{\epsfbox{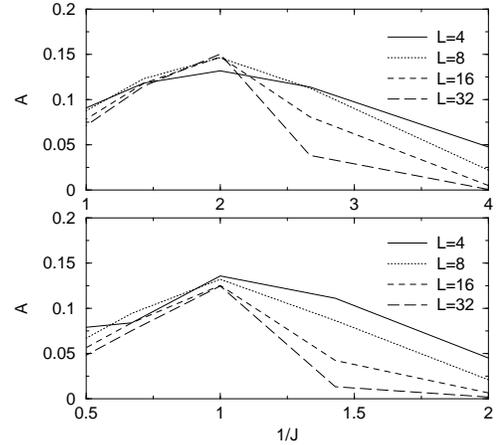}}}
  \narrowtext{\caption{$A$ as function of $1/J$ for $h=0$ and 
$L=4,8,16,32$. The top graph is for bond-moving, the bottom 
graph on a hierarchical lattice.}
\label{A_h=0}
}
\end{figure}

Next, let us discuss the quantity $A$ which is a measure of the degree
of non self-averaging. Figure \ref{A_h=0} shows our results in the
absence of a magnetic field. In the high-temperature phase as well as
in the spin-glass phase $A$ descreases with increasing system size and
approaches zero, just as one can expect in the absence of RSB.  At the
critical coupling strength $J_c$, $A$ remains constant with increasing
system size, its value being $A \simeq 0.13$ for the hierarchical
lattice, and $A\simeq 0.15$ for bondmoving. The constancy of $A$ at
the critical point can be explained from the scaling behaviour of the
overlap distribution function $P(q)$. Critical scaling implies
$$ \left[P(q)\right]=L^{-\beta/\nu} \left[\tilde 
P(qL^{\beta/\nu})\right]$$
and
$$\left[P(q)P(q')\right]=L^{-2\beta/\nu} \left[\tilde 
P(qL^{\beta/\nu})\tilde 
P(q'L^{\beta/\nu})\right] ,$$
with $\beta$ being the
order parameter critical exponent, and $\nu$ the correlation 
length
exponent. Introducing the variable $y=qL^{\beta/\nu}$, we then 
obtain
$$A=\frac{\int\int y^2 y'^2 \tilde P(y)\tilde P(y') dy 
dy'}{\left(\int y^2 \tilde P(y) dy\right)^2}-1\,$$
independently of $L$. 

For low temperatures $T=1/J$, $A$ seems to follow the law $A 
\sim
TL^{-\theta}$ with $\theta \simeq 0.24$. This can be derived 
by the
following argument: At low temperatures, most samples have a 
value of $\langle q^2 \rangle$ close to 1, and only a fraction 
$p$
proportional to $kTL^{-\theta}$ of all samples have 
system-wide
excitations  and have therefore
some other value $ \langle q^2 \rangle = x < 1$. We therefore 
obtain
$$\left[\langle q^2\rangle \right] \simeq 1-p + p [x]$$
and 
$$\left[\langle q^2\rangle^2 \right] \simeq 1-p + p [x^2],$$
leading to 
$$A \simeq p(1+[x^2]-2[x]) \sim kTL^{-\theta}.$$

In the presence of a magnetic field, we expect $A$ to decrease 
always
to zero for large system sizes, because the system is always 
in the
high-temperature phase without long-range correlations. 
However, as we
will show in the following, $A$ can become nevertheless very 
large for
certain combinations of the system size, the field, and the 
two-spin
coupling strength. One can therefore easily get the impression 
that
the system is not self-averaging, while in reality $A$ 
increases only
over a limited range of system sizes or parameters. 

\begin{figure}
  \centerline{    
\epsfysize=0.7\columnwidth{\epsfbox{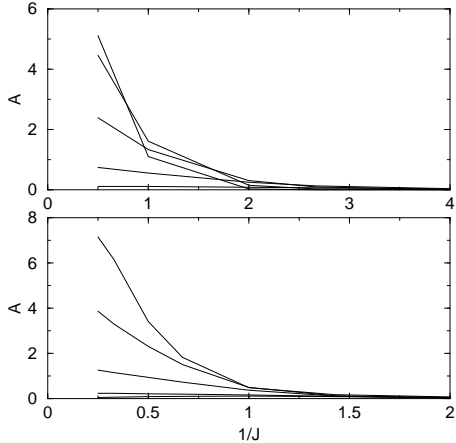}}}
  \narrowtext{\caption{$A$ as function of $1/J$ for $h=0.2$ and
$L=2,4,8,16,32$ (from bottom to top for the $1/J=0.5$ points). 
The top graph is for bond-moving, the bottom graph on a 
hierarchical lattice.}
\label{A_h=0.2}
}
\end{figure}

Figure \ref{A_h=0.2} shows our results for $A$ with a magnetic 
field
$h=0.2$.  For $J>J_c$, the values are larger than without 
field, and
they increase with increasing system size and decreasing 
temperature.
We expect that as the system size increases further, $A$ will 
reach a
maximum and then decrease again. For fields stronger than 
$h=0.5$, we
see this reversal in the trend of $A$ already for the system 
sizes
studied in the simulations. For $J<J_c$, Fig.~\ref{A_h=0.2} 
shows that
the curves for different $L$ intersect each other, such that 
for high
temperatures self-averaging is better for larger system sizes. 
Thus,
the behaviour of $A$ for weak fields seems to be qualitatively 
similar to that
of the three-spin model. 

\begin{figure}
  \centerline{
    \epsfysize=0.7\columnwidth{\epsfbox{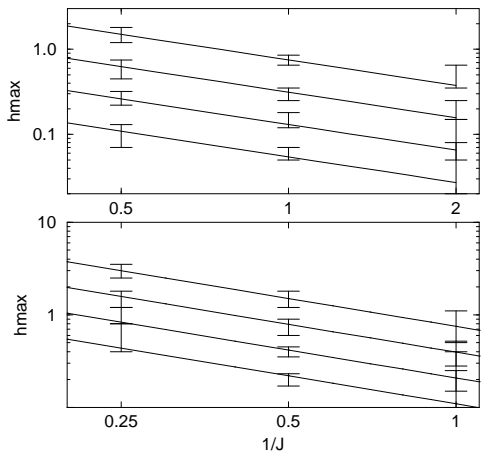}}}
  \narrowtext{\caption{The field value $h_{max}$ for which $A$ 
is largest, for $L=4,8,16,32$ (from top to bottom). The 
straight lines are power laws $h_{max}\propto JL^{-1.26}$ (top 
graph, bond-moving) and $h_{max}\propto JL^{-0.93}$ (bottom 
graph, hierarchical lattice).
}
\label{Amax}
}
\end{figure}

For given system size and coupling strength $J>J_c$, there exists
always a value of $h$ for which $A$ has a maximum. The height of this
maximum is higher for larger system sizes and for lower temperatures
$1/J$.  Figure \ref{Amax} shows the field for which $A$ is largest as
function of $1/J$. The data are in good agreement with a dependence
$h_{max} \propto J$, and $h_{max} \propto L^{-1.26}$ for bond-moving.
This means that $A$ is largest when $h \sim J L^{(d/2-\theta)}$.  For
the hierarchical lattice, the fit to the data is best for a dependence
$h_{max} \propto J$, and $h_{max} \propto L^{-0.93}$. Just as in
Figure \ref{flowmax}, the effective value of $d/2-\theta$ appears to
be larger on the hierarchical lattice than for bond-moving.  We
suspect that this is due to the fact that the field grows indefinitely
on the hierarchical lattice.

These results can be understood if one considers the effect of the
field on the overlap distribution $P(q)$. Without field, $P(q)$ is a
symmetric function, and varies considerably in shape for different
samples for the system sizes typically used in simulations.  This
feature is seen in Monte-Carlo simulations \cite{mpr98} as well as in
MKA \cite{us1}. A magnetic field changes the shape of $P(q)$ and moves
the weight more and more towards positive $q$. In the limit $h\to
\infty$, all spins are aligned with the field, leading to
$P(q)=\delta(1)$. Along the boundary line $L \sim
(J/h)^{1/(d/2-\theta)}$, where the field is not yet strong enough to
destroy all features of the low-temperature phase, we can expect that
at least some samples still have large droplets than can be flipped
without much free energy cost.  In Monte-Carlo simulations
\cite{mnz98}, one finds indeed for certain intermediate parameter
values a $[P(q)]$ that has a pronounced peak at some large $q$ value,
and a long and thin tail that extends almost all way down to $q=-1$.
The authors point out that this feature results from most samples
having only the main peak, and other samples having an additional
second peak for some other value of $q$. They go on to argue that this
is a non self-averaging feature characteristic of RSB, and that it
should not be expected if there was no RSB. However, they also admit
that their simulations do not show a second peak in $[P(q)]$ at a
value $q_{min}$, which is expected from Mean-Field Theory.  Although
we have not determined $P(q)$ in the presence of a magnetic field
within MKA, we can conclude from the behaviour of $A$ that $P(q)$ must
have in MKA approximation exactly the same features that we just
described for the Monte-Carlo simulations.  Indeed, it is easy to show 
that $A$ becomes large if most
samples have a $P(q)$ with one narrow peak at $q_0$, and some samples
have two peaks. For those samples with one peak, we have a small
variance of $q$,
$$\chi_s \equiv \langle q^2 - q_0^2 \rangle, $$
which is essentially sample-independent. For those samples 
with two peaks, we have a large variance $\chi_l$ which is 
different for each sample. If the fraction of samples with two 
peaks is $p$, we obtain
\begin{eqnarray*}
 A &=& \frac{(1-p) \chi_s^2 + p [\chi_l^2]}{((1-p) \chi_s + p 
[\chi_l])^2}-1\\
&\simeq& \frac{p [\chi_l^2]}{\chi_s^2+2p\chi_s 
[\chi_l]+p^2[\chi_l]^2},
\end{eqnarray*}
where we have only kept the leading terms. As long as $p$ is 
not much
smaller than $[\chi_l]/\chi_s$, $A$ is of the order $1/p$. 
Thus, if
$\chi_s$ is small and $p$ is small but not too small, $A$ is 
large.
The second condition is satisfied if the field is such that a 
small
fraction of samples have a second peak in $P(q)$, the first 
condition
is better satisfied for larger $L$ or smaller $T$. This 
explains why
we observe the maximum of $A$ for those $h$ values where the
contribution of the field to the free energy is comparable to 
that of
the couplings, and why the maximum of $A$ is larger for larger 
systems
and lower temperatures. Of course, for some even larger value 
of $L$,
we expect $P(q)$ to start having less sample-to-sample 
fluctuations, because
the samples should become self-averaging. Then the argument 
will break
down, and $A$ should remain small. However, this range of 
system sizes
seems to be beyond the reach of our simulations.

In conclusion, we have shown that there exists a wide range of
parameters over which the degree of non-self-averaging appears 
large
for system sizes typically used in computer simulations. We 
expect our
results to be valid beyond MKA. As we have argued for systems 
without
a field \cite{us1,us5}, the apparent non-self-averaging must 
be
attributed to the influence of the zero-field critical point. 
This
influence reaches surprisingly far and creates a line in the 
$h-J$
plane along which non-self-averaging is particularly large, 
and which
is somewhat reminiscent of the de Almeida-Thouless line.  We 
are here
in agreement with Huse and Fisher \cite{hf91} who argued 
already
almost 10 years ago that Monte-Carlo simulation data for a 
spin glass
in a magnetic field are strongly affected by the critical 
point.

\section{Discussion}

We have shown that for the three-spin model as well as for the 
spin
glass in a magnetic field, a large degree of 
non-self-averaging found
in computer simulations does not represent unequivocal 
evidence for
RSB, but can be caused by finite-size effects.

It seems, however, that a study of the non-self-averaging 
parameter
$A$ using Monte-Carlo simulations might be able to 
discriminate
between RSB and the droplet picture for three- and 
four-dimensional
spin glasses. As we have shown, $A$ has in zero field a 
maximum at
$T_c$, and decreases again with decreasing temperature in MKA. 
If
there was a low-temperature phase with RSB, the 
low-temperature value
for $A$ should probably be larger than the critical value, and 
$A$
should therefore increase with decreasing temperature. Also, 
for
temperatures below $T_c$, we found that $A$ has its maximum 
not at
zero field, but at some finite field value. The degree of
self-averaging decreases deep in the supposed low-temperature 
phase. We
expect a similar behaviour from the Monte-Carlo simulations. 
This
would be a hint that non-self-averaging is strongest along the
boundary between the field-dominated and coupling-dominated 
regime, and
not in the region where one would expect a low-temperature 
phase with RSB. 

\acknowledgements This work began when HB and BD were at the
Department of Physics, University of Manchester, supported by EPSRC
Grants GR/K79307 and GR/L38578. BD also acknowledges support from the
Minerva foundation, and from the German Science Foundation (DFG, Grant
Dr300/2-1), and she acknowledges the hospitality of the ICTP in
Trieste during a short visit there. MAM thanks Dr A. Cavagna for useful
discussions on the mean-field limit.

\end{multicols} 
\end{document}